\documentstyle[11pt,aaspp4,epsf]{article}
\received{8 June 1995}
\revised{18 Aug 1995}
\accepted{12 Sep 1995}

\newcommand{\be}{\begin{equation}}
\newcommand{\ee}{\end{equation}}
\newcommand{\bc}{\begin{center}}
\newcommand{\ec}{\end{center}}
\newcommand{\etal}{{\it et al.}}

\slugcomment{astro-ph/9509074 - UCLA-ASTRO-ELW-95-03}

\lefthead{Wright}
\righthead{Halo Primordial Black Holes}

\begin{document}

\title{On the Density of PBH's in the Galactic Halo}
\author{Edward L. Wright}
\affil{UCLA Dept. of Physics \& Astronomy}
\authoraddr{Division of Astronomy\\P.O. Box 951562\\
Los Angeles CA 90095-1562}

\begin{abstract}
Calculations of the rate of local Primordial Black Hole explosions
often assume that the PBH's can be highly concentrated into galaxies,
thereby weakening the Page-Hawking limit on the cosmological density of
PBH's.  But if the PBH's are concentrated by a factor exceeding
$c/(H_\circ R_\circ) \approx 4 \times 10^5$, where $R_\circ = 8.5$~kpc
is the scale of the Milky Way,
then the steady emission from the PBH's in the halo will produce an
anisotropic high latitude diffuse gamma ray intensity larger than
the observed anisotropy.
This provides a limit on the rate-density of evaporating PBH's of
$\lesssim 0.4$~pc$^{-3}$yr$^{-1}$ which is more than 6 orders of
magnitude lower than recent experimental limits.
However, the weak observed anisotropic high latitude diffuse gamma ray
intensity is consistent with the idea that the dark matter that closes the
Universe is Planck mass remnants of evaporated black holes.
\end{abstract}


\section{Introduction}

The average density of primordial black hole's (PBH's) in the Universe is
constrained by the Page-Hawking \markcite{PH76} (1976)
limit on the diffuse gamma ray intensity,
since the Hawking \markcite{H74} (1974) radiation from PBH's
produces copious gamma rays.
Halzen \etal\ \markcite{HZMW91} (1991)
computed the photon spectrum
from a uniform density of PBH's with the initial mass function
$n_i(m_i)dm_i \propto m_i^{-2.5}dm_i$, and found
that
$\Omega_{PBH}h^2 < (7.6 \pm 2.6) \times 10^{-9}$
with $h = H_\circ/100$~km/sec/Mpc,
or an average mass density of PBH's of
$\rho_{PBH} < 1.43 \times 10^{-37} \;\mbox{\rm gm\,\,cm}^{-3},$
and an average number density of PBH's of $N = 10^4$~pc$^{-3}$.

Now I wish to calculate the maximum allowed density of PBH's in the
halo of the Milky Way.  The halo mass density is given by
\be
\rho_h = \frac{v_c^2}{4\pi G R^2} = 8.4 \times 10^{-25}
\;\mbox{\rm gm\,\,cm}^{-3}
\ee
at the position of the solar circle, $R_\circ = 8.5$~kpc, for a circular
velocity of $v_c = 220$~km/sec.
Since the halo density is $(4.5 \times 10^{4})/\Omega h^2$
times higher than the average
density  of the Universe, one could hope that the PBH's would have a higher
density in the halo of the Milky Way which would make it easier to detect
the explosions caused by their final evaporation in high energy gamma
ray experiments.
In fact, Halzen \etal\ \markcite{HZMW91} (1991) considered concentration
factors up to
$\zeta = \rho_h/\overline{\rho} = 1.36/h^2 \times 10^{7}$
by assuming that the PBH's were as highly concentrated as the luminous
matter.  Cline \& Hong \markcite{CH92} (1992)
considered local densities as high as
$N = 10^{12}$~pc$^{-3}$ by assuming that some of the gamma-ray bursts observed
by BATSE were PBH explosions.

\section{Calculation of Anisotropy}

However, the PBH's in the halo will contribute an anisotropic diffuse
gamma ray intensity that will be much easier to measure for instrumental
reasons than the isotropic intensity.
A rough order of magnitude estimate for this anisotropic signal
is a fraction $\zeta H_\circ R_\circ/c$ times the isotropic background.
In order to improve this calculation, I need to compute the local
average emissivity from the Page-Hawking limit which is integrated over time.
For simplicity, I will do this calculation for the bolometric gamma ray
intensity, though a frequency-dependent calculation would give a better limit.
The mass spectrum of PBH's produced by a scale invariant perturbation
spectrum in the early Universe is
$n_i(m_i)dm_i \propto m_i^{-2.5}dm_i$,
and the initial mass of a PBH that is evaporating at time $t$
is $m_i \propto t^{1/3}$.
The total comoving density $\rho/(1+z)^3$ of PBH's scales like
\be
\rho/(1+z)^3 \quad \propto \quad \int_{t^{1/3}} m \; m^{-2.5} dm
\quad \propto \quad t^{-1/6}
\ee
and thus the comoving luminosity density scales like
$t^{-7/6} \propto (1+z)^{7/4}$ for $\Omega = 1$.
The bolometric intensity is related to the comoving emissivity by
\be
I = \int \frac{j_{CM}(t)}{1+z} c dt
  = \int \frac{j_\circ (1+z)^{7/4}}{(1+z)^{7/2}} \frac {c}{H_\circ} dz
\ee
This gives the relationship between the current average emissivity and the
isotropic integrated intensity,
\be
I_{iso} = \frac{4 j_\circ c}{3 H_\circ}.
\ee
Now I will calculate the emission from the halo of the Milky Way.
If the emissivity of the halo is $j_h$, then its contribution to the
anisotropic intensity at angle $\theta$ with respect to the
Galactic center is
\be
I_{aniso} = \int j_h(R) ds = \frac{\pi - \theta}{\sin\theta}
j_h(R_\circ) R_\circ
\ee
for a spherical halo with density following the singular isothermal
sphere model: $\rho \propto r^{-2}$.
This is a special case of an ``isothermal'' halo with core radius $r_c$,
flattening $q$,
and density varying like:
\be
\rho = \frac{\rho_\circ (R_\circ^2+r_c^2)}{R^2+z^2/q^2+r_c^2}
\ee
with $R$ and $z$ being cylindrical coordinates.
This gives an anisotropic intensity of
\be
I_{aniso} = \int j_h ds =
            j_h(R_\circ) R_\circ \eta(l,b,r_c/R_\circ,q)
\ee
with
\be
\eta(l,b,r_c/R_\circ,q) =
\frac{(1+(r_c/R_\circ)^2)
\left[\pi/2 + \tan^{-1}\left(\cos\theta^\prime/
\sin\theta^{\prime\prime}\right) \right]}
{\sin\theta^{\prime\prime}\sqrt{1+(q^{-2}-1)\sin^2b}}
\ee
where
$\cos\theta^\prime = (\cos l \cos b)/\sqrt{1+(q^{-2}-1)\sin^2b}$
and
$\sin\theta^{\prime\prime} =
\sqrt{1 - \cos^2\theta^{\prime} + r_c^2/R_\circ^2}.$

Most of the uncertainty in the endpoint of primordial black hole evaporation
cancels out in the ratio of $I_{aniso}/I_{iso}$ which is given by
\be
\frac{I_{aniso}}{I_{iso}} = \eta(l,b,r_c/R_\circ,q)
\frac{j_h(R_\circ)}{j_\circ}\frac{3 H_\circ R_\circ}{4 c}
\label{anirat}
\ee

\section{Comparison with Data}

\begin{figure}[t]
\plotone{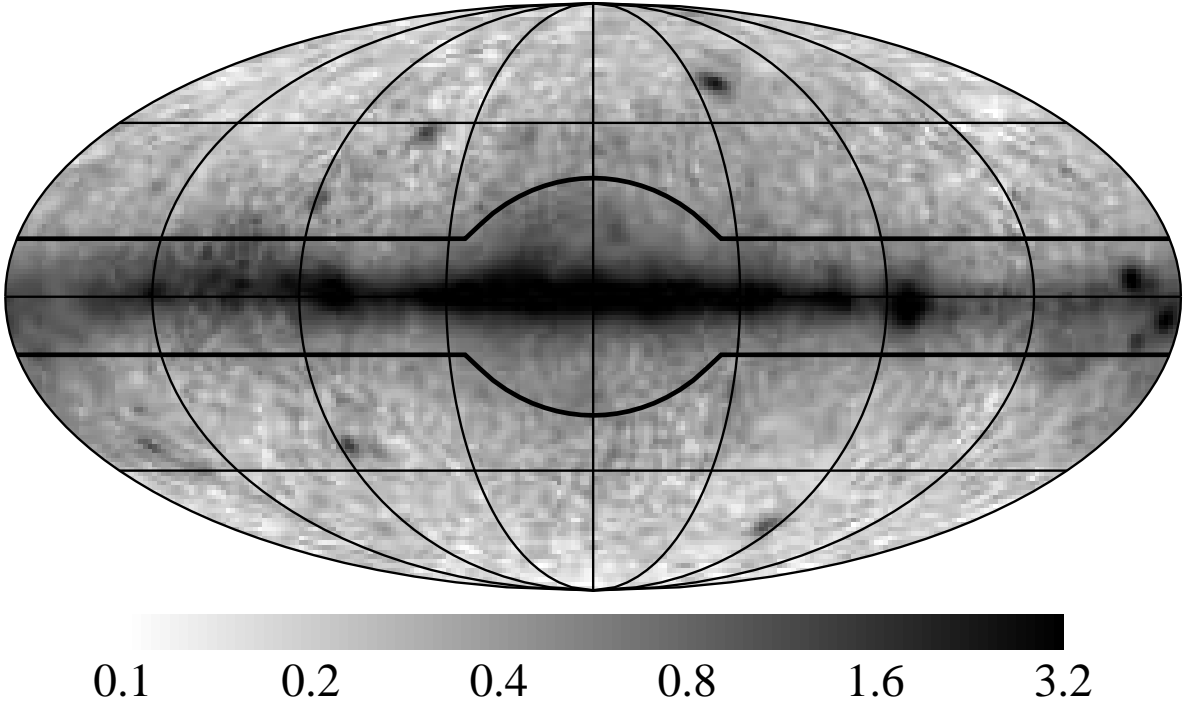}
\caption{Gamma rays with $E > 100$~MeV from EGRET.
\label{skymap.eps}}
\end{figure}

Using the combined Phase I EGRET sky maps in the directory
/compton\_data/egret/combined\_data on legacy.gsfc.nasa.gov, I have
constructed the rate map in Figure \ref{skymap.eps}.
This shows the rate of gamma rays with $E > 100$~MeV.
Figure \ref{skymap.eps} has been smoothed with a $2^\circ$ FWHM
quasi-Gaussian test function (Wright \etal\ \markcite{RMS10} 1994),
but the fits are based on unsmoothed maps.

PBH's are not the only, or even the primary, source of galactic gamma ray
emission.  The process
\be
p_{CR} + p_{ISM} \rightarrow p + p + \pi^\circ \rightarrow p + p + 2\gamma
\ee
where a cosmic ray proton $p_{CR}$ hits an interstellar medium
proton $p_{ISM}$ is the dominant source of galactic gamma rays.
Digel, Hunter \& Mukherjee \markcite{DHM95} (1995) find an average
emissivity of $(1.7 \pm 0.1) \times 10^{-26}\;$photons/sec/sr/proton
for $E_\gamma > 100$~MeV
in the Orion region which is similar to interstellar medium emission rates
elsewhere in the solar neighborhood.
While the cosmic ray density appears to be higher in the inner galaxy,
leading to a higher emissivity per ISM proton, I can avoid this complication
by not using regions close to the galactic plane in my fits.
I have used the 100 $\mu$m intensity as a proxy for the column density of
ISM protons.  This will automatically include the H$_2$ and H~II components
of the ISM that are not measured by the 21~cm neutral hydrogen line.
The 100 $\mu$m emissivity per proton will depend on the local dust
properties and radiation field, but avoidance of the galactic plane minimizes
these complications as well.
The 100 $\mu$m intensity at high $|b|$ is approximately
$0.6 \times 10^{-20}$~MJy/sr/proton/cm$^2$, so one expects about
$0.03\;$photons/m$^2$/sr with $E > 100$~MeV for each MJy/sr of 100 $\mu$m
intensity.

When I fit the gamma ray map to the form
\be
I_\gamma(l,b) = j_h(R_\circ)R_\circ\,\eta(l,b,r_c/R_\circ,q)
         + \frac{dI_\gamma}{dI_{100}} I_{100}(l,b)
         + \frac{dI_\gamma}{d\csc|b|} \csc|b|
         + I_\circ + \epsilon(l,b)
\ee
with $I_{100}$ being the 100 $\mu$m intensity from the
DIRBE instrument on {\sl COBE} with a model of the zodiacal light removed,
and $\epsilon$ being the residual,
I get the results shown in Table \ref{tab.halofit}.
Missing parameter values are fixed at their default values, which are
zero except for $r_c/R_\circ = 0.01$ and $q = 1$
(which approximate the simple singular isothermal sphere model.)
The isotropic intensity is required to be non-negative.
These fits were done using the least sum of absolute errors
$\sum|\epsilon|$ instead of
least squares fitting to avoid the effect of bright sources, and an
elliptical region surrounding the Galactic Center with $b = \pm 30^\circ$
and $l = \pm 45^\circ$ was excluded along with $|b| < 14.5^\circ$.
The excluded region is marked on Figure \ref{skymap.eps}.
The pixels used for this fit were {\sl COBE} DMR pixels with 4328 pixels
outside the exclusion region, so differences in $\sum |\epsilon|$ greater
than 0.1\% are statistically significant.
Simulating fits to 100 skies based on the best fit model in Table
\ref{tab.halofit} along with the observed residuals gave values of
$j_h(R_\circ)R_\circ$ with a standard deviation of 0.00165, which
is negligible when compared to the range of $j_h(R_\circ)R_\circ$
obtained using different galactic tracers and halo models.

The coefficient found for the $I_{100}$ term
is consistent with the result expected from studies of the interstellar
medium in the solar neighborhood (Digel \etal\ \markcite{DHM95} 1995),
but when $\csc|b|$ or flattened haloes are allowed to absorb some of the
galactic flux the coefficient is slightly lower but not unreasonable.
Fits that do not include $\csc|b|$ as a tracer tend to favor
flattened haloes with large core radii, which makes the
halo model $\eta$ approximately proportional to $\csc|b|$.
Models that do include a separate $\csc|b|$ term favor spherical,
singular isothermal sphere haloes.
Because a flattened halo has a smaller thickness at high $|b|$,
the fitted halo density is higher for the flattened models.
The large core radius in the flattened models causes the halo to
produce a large monopole contribution to the intensity, so the isotropic
term goes to zero in the flattened models.

Models similar to the large core radius spherical haloes proposed for
gamma ray burst (GRB) sources do not provide an anisotropic intensity,
for the simple reason that the GRB's are observed to be isotropic.
In these models the halo density is limited by the isotropic intensity,
and scales like $\j_h \propto r_c^{-1}$.  Thus large core radius models
are not significant for placing an upper limit on the local halo density.

\begin{deluxetable}{ccccccc}
\tablewidth{0pt}
\tablecaption{Halo fits to the EGRET map. \label{tab.halofit}}
\tablehead{
\colhead{$j_h(R_\circ)R_\circ$} &
\colhead{$dI_\gamma/dI_{100}$}  &
\colhead{$dI_\gamma/d\csc|b|$}  &
\colhead{$r_c/R_\circ$}         &
\colhead{$q^{-2}$}              &
\colhead{$I_\circ$}             &
\colhead{$\sum |\epsilon|$}
}

\startdata
\nodata & \nodata & \nodata & \nodata & \nodata & 0.2818 & 413.6 \nl
0.0454  & \nodata & \nodata & \nodata & \nodata & 0.2006 & 393.5 \nl
0.1504  & \nodata & \nodata &   2.03  &   7.5   &    0   & 303.2 \nl
\nodata & \nodata & 0.1075  & \nodata & \nodata & 0.1005 & 319.7 \nl
0.0409  & \nodata & 0.1024  & \nodata & \nodata & 0.0364 & 298.9 \nl
0.0512  & \nodata & 0.0952  &   0.01  &   1.5   & 0.0413 & 298.8 \nl
\nodata &  0.0330 & \nodata & \nodata & \nodata & 0.1692 & 310.7 \nl
0.0229  &  0.0317 & \nodata & \nodata & \nodata & 0.1345 & 303.8 \nl
0.0743  &  0.0183 & \nodata &   2.89  &   5.0   &    0   & 280.9 \nl
\nodata &  0.0218 & 0.0574  & \nodata & \nodata & 0.1099 & 289.4 \nl
0.0306  &  0.0173 & 0.0656  & \nodata & \nodata & 0.0578 & 278.1 \nl
\enddata
\tablecomments{$I_\gamma$, $I_\circ$, $j_h R_\circ$ and $\epsilon$ are
in photons/m$^2$/sec/sr,
while $I_{100}$ is in MJy/sr.}
\end{deluxetable}

Different methods of fitting for the galaxy give very different
estimates for the isotropic background, but the estimate for the
local halo density $j_h(R_\circ)$ is slightly more stable.
Thus it is more appropriate to normalize to the
Halzen \etal\ \markcite{HZMW91} )1991) model
upon which their calculation of the Page-Hawking limit is based
instead of the uncertain isotropic background.
Integrating Figure 3 of Halzen \etal\ \markcite{HZMW91} (1991)
for $E > 100$~MeV gives a flux of 0.06~photons/m$^2$/sec/sr which
is in the range of the isotropic intensities from the fits in Table
\ref{tab.halofit}, and is also consistent with the darkest sky intensity of
$0.15$~photons/m$^2$/sec/sr given by Bertsch \etal \markcite{BFHHT95} (1995).

Comparing these results with Equation \ref{anirat} implies that
\be
\frac{j_h(R_\circ)R_\circ}{(4/3)j_\circ(c/H_\circ)} =
\frac{0.023\mbox{--}0.15}{0.06} = 0.4 \mbox{--} 2.5
\ee
or
\be
\zeta = \frac{4 c} {3 H_\circ R_\circ} \times 0.4 \mbox{--} 2.5 =
(2\mbox{--}12)/h \times 10^5.
\ee
Using this $\zeta$ I get a local density of PBH's of
$N_h = \zeta \times 10^4 =
(2\mbox{--}12)/h \times 10^9 \;\mbox{\rm pc}^{-3}$.

\begin{figure}[t]
\plotone{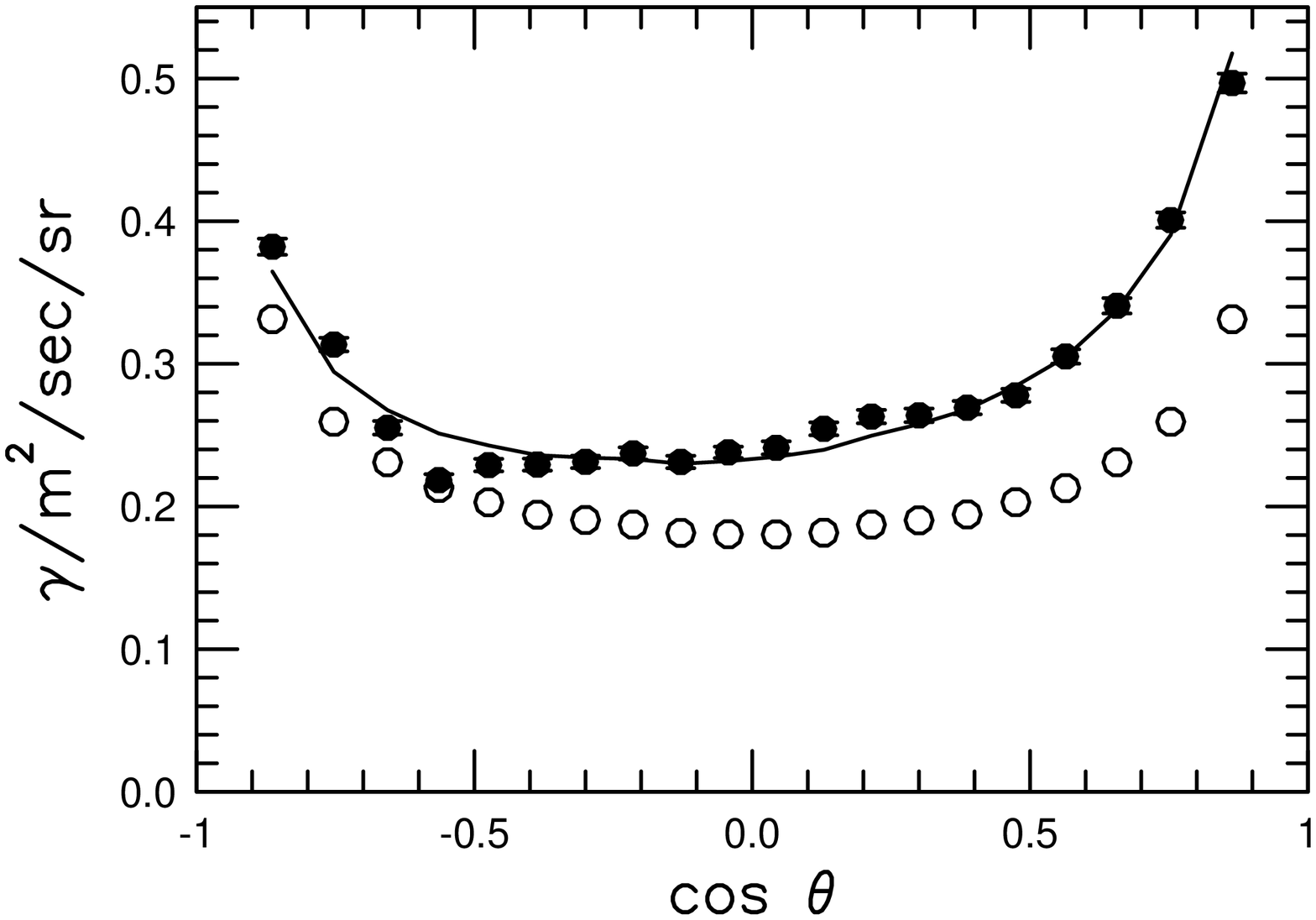}
\caption{Fit of a $\csc|b|$ plus halo model to the
$E > 100$~MeV intensity.  $\theta$ is the angle between the line-of-sight
and the galactic center. \label{ivstheta.eps}}
\end{figure}

An alternative fit is shown in Figure \ref{ivstheta.eps}.
The data with $|b| > 19.3^\circ$ was binned into twenty bins equally
spaced in $\cos\theta$.  Within each bin, the mid-average, or average
of the two middle quartiles of the data sorted by intensity, was taken.
The filled points in Figure \ref{ivstheta.eps} are these mid-averaged
intensities.
Since large $|\cos\theta|$ only occurs for small $|b|$, the mid-average
value of $\csc|b|$ was also computed in each $\theta$ bin.
The mid-averaged intensities were then fit to the form
\be
I_\gamma = I_\circ + \frac{dI_\gamma}{d\csc|b|} \csc|b| +
j_h(R_\circ) R_\circ \eta(l,b,0,0)
\ee
giving coefficients
$I_\circ = -0.005$~photons/m$^2$/sec/sr,
a slope of $dI_\gamma/d\csc|b| = 0.14$~photons/m$^2$/sec/sr,
and a local halo emissivity of
$j_h(R_\circ) R_\circ = 0.037$~photons/m$^2$/sec/sr.
This fit is the curve in Figure \ref{ivstheta.eps}, and the
contribution of the $\csc|b|$ term is shown as the open circles.

\section{Discussion}

If PBH's were strongly concentrated in the halo of the Milky Way, they
would produce a large anisotropic gamma ray flux
which could easily be observed.
A weak anisotropic signal of the predicted form is present.  With the
resulting value for the concentration factor
$\zeta \approx (2\mbox{--}12)/h \times 10^5$,
I can use the Halzen \etal\ \markcite{HZMW91} (1991)
calculation of the average PBH explosion rate to estimate that the
explosion rate of halo PBH's is
\be
\frac{dn}{dt} \leq 0.07 \mbox{--} 0.42 \; \mbox{pc}^{-3} \mbox{yr}^{-1}.
\ee
Since there are several possible emission mechanisms other than PBH's
which could be located in a galactic halo, I have taken the rate from the
fits as an upper limit.
The dominant uncertainty in this limit is the physical thickness of the
galactic density enhancement, and this is reflected in the range of models
considered in Table \ref{tab.halofit}.  The highest densities correspond to
flattened haloes, and for a collisionless species like PBH's the highest
likely flattening corresponds to an E7 galaxy shape with $q^{-2} = 11$.
The best fit values obtained here when not including a separate $\csc|b|$
term, $q^{-2} = 5$ or 7.5, are both equivalent to an E6 galaxy shape.
The observed flattening of the dark matter in polar ring galaxies is in
this range (Sackett \etal\ \markcite{SRJF94} 1994).

The uncertainties in modeling the last few seconds of PBH lifetime
have very little effect on the limit derived here, because the diffuse
gamma ray flux comes primarily from PBH's with masses $M > 10^{14}$~gm and
temperatures $kT < 100$~MeV, and radiation under these conditions is well
understood.  However, the behavior of PBH's at higher temperatures is not
so well known, and different models of the final burst can give vastly
different detection limits.  For example,
the recent EGRET limit of $dn/dt < 0.05$~pc$^{-3}$yr$^{-1}$
(Fichtel \etal\ \markcite{F94} 1994) assumed that the last
$6 \times 10^{13}$ grams of PBH rest mass evaporate producing
$10^{34}$~ergs of 100 MeV gamma rays in less than
a microsecond based on the Hagedorn \markcite{H70} (1970)
model for high energy particles,
while in the standard model of high energy particle physics
it takes $> 10^6$ years to evaporate
this mass (Halzen \etal\ \markcite{HZMW91} 1991).
The particular technique used in Fichtel \etal\ \markcite{F94} (1994)
would not be sensitive to standard model bursts.
Porter \& Weekes \markcite{PW78} (1978) derived a limit of
$dn/dt < 0.04$~pc$^{-3}$yr$^{-1}$ using the Hagedorn model,
but this limit is weakened to $dn/dt < 5 \times 10^8$~pc$^{-3}$yr$^{-1}$
in the standard model (Alexandreas \etal\ \markcite{Cygnus93} 1993).
Cline \& Hong \markcite{CH92} (1992)
proposed that an expanding fireball could convert
much of the $10^{34}$~ergs from a Hagedorn model burst into MeV gamma rays
detectable by the BATSE experiment.  Given the limit on $dn/dt$ above,
BATSE would have to be able to detect PBH explosions out to distances
${\cal O}(10^{19})$~cm if even a few percent of the BATSE bursts were
due to PBH's.
Other sensitive limits on the PBH explosion rate density (Phinney \& Taylor
\markcite{PH79} 1979) depend on conversion of the last $10^{11}$ grams
of PBH rest mass into an expanding fireball that produces a GHz radio pulse
by displacing the interstellar magnetic field (Rees \markcite{R77} 1977).
However, in the standard model it takes a few days
for the last $10^{11}$~grams to evaporate
(Halzen \etal\ \markcite{HZMW91} 1991), so no radio pulse is
generated.

Similarly, variations in the initial mass function of PBH's do not affect
the ratio of the local emissivity to the rate density of evaporating bursts,
because both the $10^{14}$~gm PBH's radiating the diffuse gamma rays
and the $10^{9}$~gm evaporating PBH's were initially formed with very similar
masses near $10^{15}$~gm.  Thus changing the slope of the initial mass
function has very little effect on the ratio of their abundances, even
though such a slope change has a large effect on the Page-Hawking limit on
$\Omega_{PBH}$.

The limit derived in the paper
on the local rate-density of evaporating PBH's provides a very
difficult target for all techniques to
directly detect standard model PBH explosions.
For example, the CYGNUS experiment presented a limit
$dn/dt < 8.5 \times 10^5$~pc$^{-3}$yr$^{-1}$
(Alexandreas \etal\
\markcite{Cygnus93} 1993), and my estimate is 6--7 orders of magnitude
smaller.

On the other hand, the ratio of the halo density of PBH's to
the Page-Hawking limit is quite close the ratio of the total halo
density to the critical density in the Universe.  This suggests
that PBH's could be tracers of the Cold Dark Matter (CDM).
This would naturally occur if the PBH's {\em were} the CDM,
but this requires either a modified mass spectrum with an enhanced abundance
of PBH's with $M > 10^{17}$~gm, or else that evaporating PBH's leave behind
a stable Planck mass remnant (MacGibbon \markcite{M87} 1987;
Carr, Gilbert \& Lidsey \markcite{CGL94} 1994).
In general, the fits in this paper give concentration ratios that are
slightly too high for this hypothesis if $\Omega = 1$ and
$h = 0.8$.  The lowest ratio of
$j_h(R_\circ)R_\circ$ to $I_\circ$ in Table \ref{tab.halofit} gives
a concentration of $\zeta_{low} = 8/h \times 10^4$, while if PBH's trace
the dark matter the expected ratio is $(4.5 \times 10^{4})/\Omega h^2$.
These are equal only if $\Omega h = 0.54 \zeta_{low}/\zeta$.  For a more
typical low value of $\zeta \approx 2/h \times 10^5$
(corresponding to $dn/dt \approx 0.1$~pc$^{-3}$yr$^{-1}$),
the required value of $\Omega h$ is $\approx 0.2$, which is consistent
with theories of large-scale structure formation in CDM
(Peacock \& Dodds \markcite{PD94} 1994)
or CDM+$\Lambda$
(Efstathiou, Sutherland \& Maddox \markcite{ESM90} 1990) models.
If this admittedly weak correspondence is correct, then 100 MeV gamma rays
are providing the first non-gravitational evidence for CDM.
Any model for dark matter that gives a 100 MeV gamma ray emissivity
proportional to the density, such as particles with a
very slow radiative decay, would also be supported by this correspondence,
while models with emissivity proportional to $\rho^2$, such as
annihilating particles, would not.
More data and better galactic modeling are needed to test
this exciting possibility.

One possible test is to look for gamma rays from large concentrations of
dark matter.
The flux from the Galaxy, $F = \int I \cos\theta d\Omega$, in
Figure \ref{skymap.eps} is $F = 1.1\;\gamma/\mbox{m}^2/\mbox{sec}$
while the flux from the halo term of the best fitting model
in Table \ref{tab.halofit} is $F = 0.38\;\gamma/\mbox{m}^2/\mbox{sec}$.
Scaling the latter flux to the mass and distance of M87
(Stewart \etal\ \markcite{SCFN84} 1984),
assuming a constant gamma ray luminosity to mass ratio, predicts a flux of
$4 \times 10^{-5}\;\gamma/\mbox{m}^2/\mbox{sec}$ for $E > 100$~MeV
which is only 10 times lower than the limit of
$40 \times 10^{-5}\;\gamma/\mbox{m}^2/\mbox{sec}$ reported by
Sreekumar \etal\ \markcite{SBDEF94} (1994).
Detection of a gamma ray flux from clusters of galaxies that correlates
with dark matter column density instead of gas column density would
support of association of dark matter and PBH's.

\acknowledgements
This research has made use of data obtained through the Compton Observatory
Science Support Center GOF account, provided by the NASA-Goddard Space Flight
Center.

\end{document}